\documentclass[conference]{IEEEtran}

\usepackage[T1]{fontenc}% optional T1 font encoding
\usepackage{mathtools}
\usepackage[inline]{enumitem}

\usepackage{mathtools}

\usepackage{graphicx}
\usepackage[tight,footnotesize]{subfigure}
\usepackage{epstopdf}
\graphicspath{{Figures/}}

\usepackage{url}
\usepackage{multirow}
\usepackage{balance}
\usepackage[numbers,sort&compress]{natbib}
\usepackage{amsmath}
\usepackage{amssymb}
\usepackage{bm}
\usepackage{lipsum}
\usepackage{color}
\usepackage[table,xcdraw]{xcolor}
\usepackage{threeparttable}

\usepackage{physics} % for abs

\usepackage{multirow}
\usepackage{authblk}

\usepackage[linesnumbered,ruled,vlined]{algorithm2e}
\SetCommentSty{mycommfont}

\SetKwInput{KwInput}{Input}                % Set the Input
\SetKwInput{KwOutput}{Output}              % set the Output

\renewcommand{\baselinestretch}{.980}

\usepackage{lscape}
\usepackage{longtable}

\begin{document}

\title{Multi-Objective Deep Reinforcement Learning for 5G Base Station Placement to Support Localisation for Future Sustainable Traffic}
%\title{Deep Learning-based Fingerprinting for Outdoor UE Positioning in 5G Networks}

\author[1]{Ahmed Al-Tahmeesschi}
\author[2]{Jukka Talvitie}
\author[3,4]{Miguel L\'{o}pez-Ben\'{i}tez}
\author[1]{Hamed Ahmadi}
\author[5]{Laura Ruotsalainen}
\affil[1]{School of Physics Engineering and Technology, University of York, United Kingdom}
\affil[2]{Unit of Electrical Engineering, Tampere University, Tampere, Finland}
\affil[3]{Dept. of Electrical Engineering and Electronics, University of Liverpool, United Kingdom}
\affil[4]{ARIES Research Centre, Antonio de Nebrija University, Spain}
\affil[5]{Dept. of Computer Science, University of Helsinki, Helsinki, Finland}
\IEEEoverridecommandlockouts
% make the title area
\maketitle
\IEEEpubidadjcol

\begin{abstract}
Millimeter-wave (mmWave) is a key enabler for next-generation transportation systems. However, in an urban city scenario, mmWave is highly susceptible to blockages and shadowing. Therefore, base station (BS) placement is a crucial task in the infrastructure design where coverage requirements need to be met while simultaneously supporting localisation. This work assumes a pre-deployed BS and another BS is required to be added to support both localisation accuracy and coverage rate in an urban city scenario. To solve this complex multi-objective optimisation problem, we utilise deep reinforcement learning (DRL). Concretely, this work proposes: 1) a three-layered grid for state representation as the input of the DRL, which enables it to adapt to the changes in the wireless environment represented by changing the position of the pre-deployed BS, and 2) the design of a suitable reward function for the DRL agent to solve the multi-objective problem. Numerical analysis shows that the proposed deep Q-network (DQN) model can learn/adapt from the complex radio environment represented by the terrain map and provides the same/similar solution to the exhaustive search, which is used as a benchmark. In addition, we show that an exclusive optimisation of coverage rate does not result in improved localisation accuracy, and thus there is a trade-off between the two solutions.

%(my old) In addition, we show that optimising for coverage rate only does not mean improving the localisation accuracy and there is a trade-off between the two solutions.
\end{abstract}

\IEEEpeerreviewmaketitle

\vspace{-2mm}

\section{Introduction}
\label{sec:Introduction}
% template source and instructions
% https://www.cambridge.org/core/journals/journal-of-navigation/information/author-instructions/preparing-your-materials

%% Localisation for AV
%Localization requirements for autonomous vehicles => provided my 5G mmWave => Infrastructure => for it to be sustainable (aesthetic and power consumption) the infra must be optimized.  

%Electrification of road transport is the primary technological change needed to meet the carbon reduction targets. However, electrification is unlikely to be sufficient since the electricity production will not be carbon neutral in the near future. There is a second major technological transformation on-going in road transport—digitalisation—bringing forth the advent of connected automated vehicles (AV). 
% major technological transformation on-going in road transport—digitalisation—bringing forth the advent of connected automated vehicles (AV). 
At present, road transport contributes a significant amount to the total carbon dioxide (CO2) emissions in the EU \citep{co2_emmission}. Thus, cities are looking for practical strategies to make their transport systems more intelligent, efficient and sustainable. One promising solution is in the form of connected automated vehicles (AV). Next-generation transportation systems represented by AVs will require Vehicle-to-Infrastructure (V2I) wireless connectivity \citep{V2I_2010}. Such connection should be able to satisfy the required high-data rates, low latency and decimeter localisation accuracy \citep{3GPP_positioning_requirments}.

5G networks have already adopted millimetre wave (mmWave) along with massive multiple-input-multiple-output (MIMO) technologies to provide extremely high data rates, low latency and localisation. However, due to unfavourable propagation conditions at high frequencies, mmWave signals experience higher path-loss and are more susceptible to building blockages than sub-6 GHz bands in urban scenarios \citep{mmWave_beamforming, wijdan_survey}. Therefore, careful planning of the base station (BS) locations is essential to reduce infrastructure costs while maintaining the quality of service and localisation accuracy \citep{cellular_localisation_survey}.

%mmWave-V2I links is capable of providing gigabit-per-second data rates and ultra-low latency.

%For instance, the 3rd Generation Partnership Project (3GPP) has set a tight positioning requirements for AV of one decimeter \citep{3GPP_positioning_requirments}. 

%%%%%%   About RL and applicaiton to wireless
Reinforcement learning (RL) is a promising technique that can be employed to address this problem. RL deals with sequential decision-making problems. The goal of a sequential decision-making problem is to select actions to maximize long-term rewards \citep{SuttonRL}. RL and the deep RL (DRL) variant have been used in the literature to optimise various wireless communications systems, for example, relay nodes selection \citep{kim-vtt-rl-nodes} and dynamic spectrum access channel selection \citep{DRL_DSA_2019}. More details on RL application to wireless communications can be found in \citep{Survey_RL_iot_2021}. In this work, we propose the utilisation of DRL to jointly optimise the coverage rate and localisation accuracy. 
% and power allocation and resource management \citep{RL_power_allocation_2019}

%%%%% general discussion about BS placement
% Elaborate more on the difference 

The problem of BS location optimisation has already been addressed by 
several studies \citep{IEEE_letter_2016, vtc_2019, Jie_cell_planning} utilising genetic algorithms or computational geometry combined with optimisation tools. 
%or to increase the line of sight probability
In addition, DRL algorithms have been mainly used for aerial BS placement to operate alone or to support terrestrial network infrastructure to improve users coverage and throughput. For instance, a single deep Q-network (DQN) agent is used to control either a single aerial BS \citep{RL_ABS_2019_IFIP, RL_ABS_2022_IoTJ} or multiple aerial BSs \citep{RL_ABS_2021_MDPI}. 
A multi-agent RL (MARL) approach is utilised to control multiple aerial BSs \citep{RL_ABS_2020_ICC}.
%% BS optimisation with RL
% DRL have been used to optimise Aerial BSs, This differnet by solving the joint optimisation problem
In the aforementioned works, the DRL has been mainly used for aerial BS location placement, while our work considers street level base stations. In addition, our work assumes a pre-deployed BS in the service region, and the proposed algorithm is capable to adapt for changes in the pre-deployed BS location.
%However, all the mentioned research mainly focuses on maximising the coverage rate for a given urban area. While this work considers the joint problem of maximising the coverage rate and minimising the localisation accuracy for users operating in an urban area scenario.
The contributions of this work are outlined as follows:
\begin{enumerate}		
	\item We investigate the BS placement with the objective of jointly optimising the coverage rate and localisation accuracy, particularly in the presence of a pre-deployed BS. This addresses the challenge of achieving both coverage requirements and accurate localisation in urban city scenarios using mmWave technology.
 
	\item We propose a solution based on DQN to tackle the multi-objective problem. The DQN framework incorporates a novel state representation approach, using a three-layered grid, which enhances the adaptation to the dynamic radio environment. Additionally, we design a suitable reward function to guide the DQN agent towards finding solutions that balance between coverage rate and localisation accuracy.

	\item We demonstrate the effectiveness the DQN framework in adapting to changes in the radio environment, as represented by the repositioning of the pre-deployed BS. The DQN model, utilising the proposed state representation approach, showcases the capability to learn and adjust in complex radio environments.	
\end{enumerate}

%The remainder of this paper is organised as follows. First, Section \ref{sec:Network model} presents the generation of the 5G mmWave signals and the considered city layout. Section \ref{sec:Problem formulation} discusses the BS placement optimisation problem and typical algorithms to solve it. Section \ref{sec:Placement optimisation with DRL} provides a brief description of the DRL algorithm along with proposed states representation, reward function shaping and implementation details. The performance of the considered models is evaluated and discussed in Section \ref{sec:Performance analysis}. Finally, Section \ref{sec:Conclusions} summarises and concludes this work.

%%%%%%%%%%%%%%%%%%%%%%%%%%%%%%%%%%%%%%%%%%%%%%%%%%%%%%%%%%%%%%%%%%%%%
 \vspace{-1mm}

\section{5G new radio network model}
\label{sec:Network model}
 \vspace{-1mm}

In this section, the 5G radio network model is described. In order to generate a realistic wireless simulation environment, the mmWave signals are generated using a ray-tracing-based approach, as recommended by 3GPP \citep{3GPP_Channel_model}. Furthermore, as our city model we select the Madrid grid, developed by the METIS project \citep{metis2020}, to represent a generic European city layout. The realizations of the ray-tracing-based radio channel are evaluated by using Wireless InSite\textregistered software~\citep{remcom}. A similar environment has been considered in the literature \citep{s20247124, icl-gnss-2022}. 

\begin{figure}[!t]

	\centering
	\includegraphics[clip, trim=0.0cm 0cm 0.0cm 0.3cm, width=0.95\columnwidth]{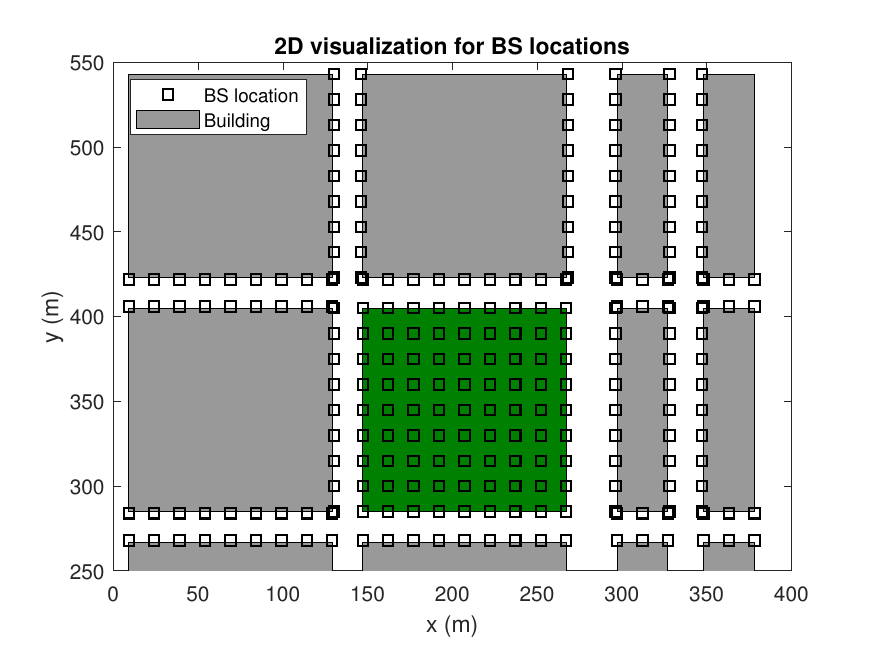}
  \vspace{-4mm}
	\caption{Madrid grid segment with candidate BS locations.}
	\label{fig:Urban-area}
 \vspace{-2mm}
\end{figure}

For this work, a specific segment of the Madrid grid is selected with BSs operating at 28 GHz. The BSs height are set to 9 m and each BS includes 4 sectors, where each sector includes a uniform linear array with 32 half-wavelength-separated patch antenna elements. The azimuth orientations of the sectors are set to 45$^\circ$, 135$^\circ$, 225$^\circ$ and 315$^\circ$. The utilised beamforming technique is implemented following the phased-array principle with a total of 64 beams per BS. The transmit power is set to 10 dBm. Fig. \ref{fig:Urban-area} shows the considered Madrid grid segment along with the candidate BSs locations. In this work, the received signal strengths (RSSs) are utilised for the estimation of the area coverage rate and localisation accuracy. In practice, the RSS measurements are obtained by the UE based on 3GPP-specified synchronisation signal blocks (SSBs) transmitted by each BS over the 64 beams \citep{3GPP_SSB}. One of the clear benefits of the considered approach is that the SSBs are continuously available in all 5G NR networks as part of standard network operation, and thus there is no need for dedicated positioning reference signals whose availability can considerably vary in practical deployments. Moreover, RSS measurements are  preferred, as they are available in the user device during both the connected mode and the idle mode as part of underlying mobility management procedures.
%Furthermore, the ray-tracing-based channel measurements are obtained by Wireless InSite software \citep{remcom}. 
%A similar simulation environment setup can also be found in \citep{s20247124, icl-gnss-2022}.
%For this work, a specific segment of the Madrid grid is selected with BSs operating at 28GHz. The BS height is set to 9m and each BS includes four sectors. Each sector includes a patch antenna element arranged in a uniform linear array. The transmitted power per sector is set to 10dBm. %(30dBm in prev paper)

%In practice, the RSS measurements obtained by the UE are based on 3GPP-specified synchronization signal blocks (SSBs) transmitted by each BS over 64 beams \citep{3GPP_SSB}. The utilised beamforming is implemented following the phased-array principle. Each of the two sectors found in a single BS is designated to cover 90 degrees and the 16 beams set for each sector are uniformly spaced (i.e., each AP covers 360 degrees). Fig. \ref{fig:Urban-area} shows the considered Madrid grid segment along with the candidate BSs locations. In this work, A single BS is assumed to be already deployed and another BS need to be added to support both user localisation and coverage.
%The detailed simulation parameters used for the simulation are provided in Table \ref{tab:sys params} and 

 \vspace{-1mm}
\section{Problem formulation}
\label{sec:Problem formulation}
 \vspace{-1mm}

Knowing the system setup, we present the BS placement strategy to jointly optimise the localisation accuracy and the coverage rate. After that we also discuss traditional exhaustive search algorithms.
%In this section, the BS placement to jointly optimise the localisation accuracy and the coverage rate is presented. After that we also discuss traditional exhaustive search algorithms.

 \vspace{-2mm}

\subsection{Optimisation problem}
We have a multi-objective optimisation problem where we need to find the optimal location of the BSs to minimise the average error in localisation and maximise the coverage rate. The optimisation problem can be formulated as:
%(it will be easier to mainly target the average localistion error $\bar{E}$ at first) 
% max f1 ∧ min f2
% max f1(x,y) AND min f2(x,y)
\vspace{-1mm}
\begin{equation}
\label{optimisation problem}
\begin{aligned}
\max\{f_1(x,y)\} \hspace{1mm} \land   &  \hspace{1mm} \min  \{f_2(x,y)\}\\
\textrm{s.t.} \quad & x \in X \\
  & y \in Y     \\
\end{aligned}
\vspace{-1mm}
\end{equation}
where $X$ and $Y$ represent the sets of potential coordinates for the BS, 
$f_1(x,y)$ is the average coverage rate (in percent) of the area and $f_2(x,y)$ is the average localisation error (in meters). The average coverage rate $f_1(x,y)$ is given as
\vspace{-1mm}
\begin{equation}
	\label{eq:f1-1}
	 f_1(x,y) =  \dfrac{1}{N} \sum_{n=1}^{N}{C_n},
  \vspace{-1mm}
\end{equation}
where $N$ is the number of positions to be covered and $C_n$ is the coverage rate and is given as
\vspace{-1mm}
\begin{equation}
	\label{eq:f1-2}
	 C_n = \begin{cases}
          1 \text{  if RSS  $\geq $ $\delta$,}
        \\
        0 \text{    otherwise}.
  \end{cases}
  \vspace{-1mm}
\end{equation}
where RSS is the received signal strength from one of the BS beams and $\delta$ is the threshold for the minimum power required for a correct signal reception. The average localisation error $f_2(x,y)$ is computed as
%minimum power received from one of the BS beams
\vspace{-1mm}
\begin{equation}
	\label{eq:f2-1}
	 f_2(x,y) = \dfrac{1}{N} \sum_{n=1}^{N}{z_n},
  \vspace{-1mm}
\end{equation}
where $z_n$ is the Euclidean distance between the estimated user equipment (UE) horizontal plane position and the actual position and is given as
\vspace{-1mm}
\begin{equation}
	\label{eq:f2-2}
	z_n  = \sqrt{({x_n^{ue}} - \hat{x}_n^{ue})^2 + (y_n^{ue} - \hat{y}_n^{ue})^2},
 \vspace{-1mm}
\end{equation}
where $x_n^{ue}$ and $y_n^{ue}$ are the user x-coordinate and y-coordinate, respectively. Moreover $\hat{x}_n^{ue}$ and $\hat{y}_n^{ue}$ are the estimated x-coordinate and y-coordinate of the user, respectively.

For the estimation of the user position, fingerprinting with the traditional K-nearest neighbour algorithm (KNN) is considered as it is one of the most utilised algorithms for RSS-based fingerprinting \citep{KNN_most_used_fingerprinting}. In KNN, the position of the UE is estimated based on the mean of $K$ nearest neighbours locations. The Euclidean distance is used to find the nearest neighbours from the accumulated database. After a brief optimisation of localization performance, in this work, we have defined $K=2$. For more details on the $K$ value estimation, please refer to \citep{icl-gnss-2022}. 
% The UE location is estimated utilising 2-KNN distance.

 \begin{figure}[!t]
	\centering
	\includegraphics[clip, trim=6.0cm 3.5cm 0.0cm 3.1cm,width=1.25\columnwidth]{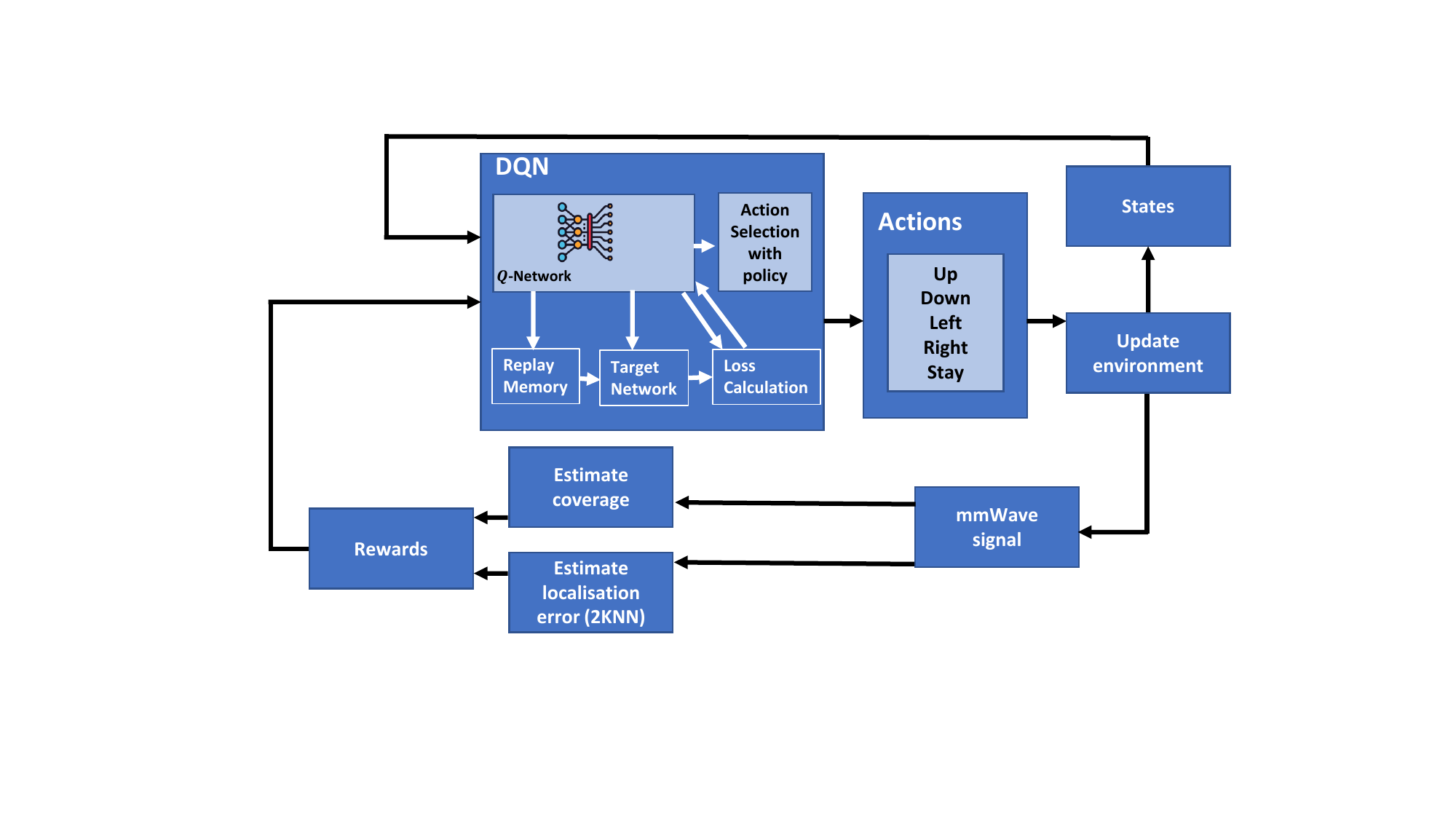}
	\caption{BS placement optimisation with DQN.}
	\label{fig:flow_graph}
 \vspace{-4mm}
\end{figure}

%Notes
%\begin{itemize}
%\item Equal weights are given for $f_1(x,y)$ and $f_2(x,y)$.
%\item Only the average localisation error is considered (i.e., without %the variance). If the variance to be included, big question is how to %include it in the rewards function.
%\item The current DQN solution for the BS position is a trade-off %between coverage and localisation accuracy
%\end{itemize}
\subsection{Brute force algorithm (BF)}
A brute force (BF) algorithm performs an exhaustive search through all the possible solutions and selects the solution that provides the best answer for the given optimisation problem. In this work, three different BF approaches are considered as a reference:
%are used to assess the performance of the DQN algorithm. 
\begin{enumerate}		
	\item BF coverage (BFC), to maximise the coverage rate (i.e., max $f_1(x,y)$) for the given area of interest.
	\item BF localisation (BFL), selects the solution that minimises the localisation error (i.e., min $f_2(x,y)$) for the area of interest.  
	\item BF joint (BFJ), selects the solution that maximises the coverage rate and minimises localisation error for the area of interest. There are multiple ways to address the general optimisation problem in (\ref{optimisation problem}). In this work, we chose to maximise the ratio $f_1(x,y)/f_2(x,y)$ as both $f_1(x,y)$ and $f_2(x,y)$ have different numerical ranges \citep{Emil_moop_2014}.
\end{enumerate}	

These BF approaches are considered in this work in order to provide a point of reference to which the performance of the proposed DQN algorithm can be compared. This is possible due to the discretisation of the space of candidate BS locations.
%In general, BF approach, the only reason to implement them in our system is for comparison reasons and due to the discretisation of the simulation endearment and the possible placement locations of the new BS.

\section{Placement optimisation with DRL}
\label{sec:Placement optimisation with DRL}
In this section, we introduce the DRL model, which incorporates the proposed state representations and reward signal shaping, aimed at optimising the BS placement for the given multi-objective problem. This problem involves jointly optimising the coverage rate and localisation accuracy, under the condition that one BS has already been deployed. This represents a simplified and tractable version of a scenario where some BSs may have already been deployed, perhaps only bearing in mind the coverage rate, and new BSs are added to also include the localisation accuracy as a relevant aspect in the infrastructure design.

%  \begin{figure*}[!t]
% 	%\vspace{-2mm}
% 	\centering
% 	\includegraphics[clip, trim=0.0cm 3.5cm 0.0cm 3cm,width=1.25\columnwidth]{flow_graph.pdf}
% 	\caption{BS placement optimisation with DQN.}
% 	\label{fig:flow_graph}
% \end{figure*}

%%%%%%%%%%%%%%%%%%%%%%%%%%%%%%%%%%%%%%%%%%%

\subsection{DRL Algorithm}
In RL, the agent continuously learns according to the rewards or punishments obtained from interacting with the environment. The agent aims to optimise the location of BSs based on coverage area and localisation accuracy. At each time step $t$, the agent observes the state $s_t$, executes an action $a_t$ following a policy $\pi$, then receives instant reward
$r_t$, and transits to the next state $s_{t+1}$, which together form a sequence
$(s_t, a_t, r_t, s_{t+1})$ of Markov Decision Process (MDP) \citep{SuttonRL}. 

The goal of an MDP is to find the optimum policy that maximises the long term discounted rewards, given by
\vspace{-1mm}
\begin{equation}
G_t = r_t + \gamma r_{t+1} + \gamma^2 r_{t+2} + \dots = \sum_{l=1}^{\infty}{\gamma^{l} r_{t+l}},
\vspace{-1mm}
\end{equation}
where $\gamma \in [0,1)$ denotes the discount factor for weighting the future rewards. If $\gamma$ is close to 0, the RL agent will focus on actions that maximise the short-term rewards, whereas if $\gamma$ is close to 1, the agent favours actions with long-term rewards. The policy $\pi$ is defined as the mapping from states to actions. In Q-learning \citep{SuttonRL}, the agent optimises the policy to maximise the action-value function $Q$, which is the expectation of the rewards $G_t$ at the current state $s_t$ and action $a_t$ under policy $\pi$ and can be described as 
\vspace{-1mm}
\begin{equation}
Q(s,a|\pi) = E_{\pi} [G_t|s_t=s, a_t=a].
\vspace{-1mm}
\end{equation}

%In Q-learning \citep{SuttonRL}, the agent determines the optimal policy $\pi$ (i.e., $\pi^\ast =$ arg max$_{a}$ $Q^\ast (s_t, a_t)$ to maximise the state-action value function $Q$ to maximise the long term rewards.

%The basic form of the Q-learning algorithm is given as: 
%\begin{equation}
%Q(s_t, a_t) \leftarrow Q(s_t, a_t) + \alpha (R_{t+1} + \gamma max Q(s_{t+1}, a_t - Q(s_t, a_t),
%\end{equation}

However, traditional RL can only work with problems that have a limited number of states and actions, which is not applicable in our case. Therefore, deep neural network is used as an approximator to the $Q$ function.
In addition, we apply experience replay which samples the data offline, prevents \textit{catastrophic forgetting} and utilises the target network. At the start of training, the target network is identical to the Q-network. As training progresses, the target network's parameters are updated less frequently than those of the Q-network. This approach is adopted to provide stability to the learning process \citep{DRL_inAction_book2020}.
% , resulting in a lag behind the Q-network's parameters

%where $Q(s_t, a_t)$ is the action value function and $s_t, a_t$ are the current states and actions, %respectively.  

The DQN is trained to minimise the loss function given as
\vspace{-1mm}
\begin{equation}
\label{eq:loss_function}
L(\theta) = \mathbb{E} [(y_t - Q(s_t, a_t ; \theta))^{2} ],
\vspace{-1mm}
\end{equation}
where $\theta$ is a vector and represents the DQN weights that determines the policy $\pi$. The target function $y_t$ is given as
\vspace{-1mm}
\begin{equation}
y_t = r_{t} + \gamma \underset{a}{\text{max}}\ Q(s_{t+1}, a_t ; \theta_{\text{target}}),
\vspace{-1mm}
\end{equation}
where $\theta_{\text{target}}$ is the target network weights and is copied from $\theta$ every fixed number steps.
%For the  case, the target action-value is found from

\subsection{DRL Agent and actions space}
In the proposed DRL algorithm, the agent is the BS (we will refer to the agent BS as ABS) who is trying to find the best deployment position to improve both localisation accuracy and coverage rate given that a BS has already been deployed. The agent's actions controls the location of the ABS. Therefore, we have 5 actions taken from discrete space $A = \{up, down, left, right, stay\}$, which represents moving the current ABS location to one of the directions (in the 2D domain) or to keep the ABS in the same position.

%In the proposed algorithm, the agent is the BS (we will refer to it as ABS) whose trying to find the best deployment position to improve both localisation accuracy and coverage rate given that a BS has already been deployed, actions and rewards are as follows:
\subsection{Proposed state representation}
% Having buildings in the considered area as the signal propagation will depend area infrastructure (building locations).
%two reasons.  The second reason is this approach can not
A typical approach for the DRL state is to have the coordinates of the pre-deployed BS and ABS as the input state. However, such approach is not suitable as it cannot adapt to changes in the starting location of the pre-deployed BS and a new training should be performed to incorporate the change in the wireless environment. Therefore, we aim to have a single algorithm that is capable to adapt for the changes in the location of the pre-deployed BS and capturing the signal propagation characteristics due to building shadowing. In the context of our work, we propose the state to be represented as a three-layered grid. The first layer represents the location of buildings. The second layer represents the location of pre-deployed BS and the third layer represents the location of the ABS. Thus, for the considered Madrid grid segment, the states are represented by a 3$\times$19$\times$24 tensor (i.e., composed of 3 layers of a 19$\times$24 grid). The grid resolution could be varied depending on the separation between the BS candidate positions, in our case 19$\times$24 is a good trade-off between complexity and performance. Each grid layer represents the position of each individual object and defined as
\begin{itemize}
\item Building layer:  1 for buildings, 0 otherwise.
\item Static BS layer: 1 for already deployed BSs, 0 otherwise.
\item Agent BS layer:  1 for agent BS, 0 otherwise.
\end{itemize}  

Algorithm \ref{alg_states} summarises the states extraction process.

% https://tex.stackexchange.com/questions/62720/vertical-space-after-algorithm
%\begin{figure}[t]
\begin{algorithm}[t]
\DontPrintSemicolon
\KwInput{Pre-deployed BS and ABS locations}
\KwInput{The Madrid grid city layout.}
\KwOutput{Tensor 3 $\times$ x $\times$ y.}

Construct the 2D grid ($x \times y$)\\

Construct the Building layer: give 1 for buildings, 0 otherwise.\\

Construct the Pre-deployed BS layer: give 1 for already deployed BSs, 0 otherwise.\\

Construct the Agent BS layer: give 1 for agent BS, 0 otherwise.\\

\caption{Obtaining the states from city layout}
\label{alg_states}
\end{algorithm}
\vspace{-4mm}

%\end{figure}

  %$S = (BSx_1, BSy_1, BSx_2, BSy_2, \ldots, BSx_m, BSy_m)$, where $m$ is the total number of BSs in the service area. A straight forward assumption is to have the location of BSs in x and y coordinates as the states for the network (this approach is used in several papers). 
  
\subsection{Reward signal shaping}
In our scenario, where the objective is to maximise the coverage rate and minimise the localisation error, it is important that the reward $r_t$ at each time step $t$  reflects the joint optimisation problem. To achieve this, we have chosen to maximise the ratio $f_1(x,y)/f_2(x,y)$. Therefore, our reward function $r_t$ is a scalar and is given as 
\begin{equation}
\vspace{-1mm}
r_t = \dfrac{f_1(x,y)}{f_2(x,y)} + p,
\vspace{-1mm}
\end{equation}
where $f_1(x,y)$ represents the coverage area percentage obtained from (\ref{eq:f1-1}) and $f_2(x,y)$ represents the localisation error obtained from (\ref{eq:f2-1}). The term $p$ corresponds to a penalty to discourage illegal actions taken by the agent, such as colliding with a building or moving outside the simulation environment. To reduce the reward when an illegal action is chosen we set $p$=-0.1, and define $p$=0 otherwise.

%In our scenario of jointly optmising for coverage rate and localisation accuracy, thus the at each time step $t$ the rewards function should maximise. For this work, we chose to maximise the ratio $f_1(x,y)/f_2(x,y)$. While it is also possible to utilise other rewards function such as to maximise . 
%it is also possible to follow $f_1(x,y) - f_2(x,y)$
%In order to combine them, we chose to optimise the ratio 
%In addition, a and b components are scaled to have values between [0,1].
%reflect increasing the ratio
%For simplicity, in this work the DQN controls the placement of 2 BSs.

\subsection{Proposed DQN training and application}

The training for the proposed DQN algorithm for the ABS placement is given by Algorithm \ref{alg_DQN_training}.
The algorithm starts by initialising the parameters (Lines 1$\sim$3), followed by $M$ episodes. Each episode includes $T$ steps and starts with resetting the states (Line 5).
For the actions selection (Lines 7 and 8), we utilise the $\epsilon$-greedy approach to balance exploration and exploitation from previous experience. After an action $a_t$ is performed a reward $r_t$ is received and transit to a new state $s_{t+1}$. The transition tuple $\{s_t, a_t, r_t, s_{t+1} \}$ is stored in experience replay memory $D$, which stores experiences in a first-in-first-out manner (Line 10). Once $D$ is larger than the mini-batch size, the network training starts (Lines 11$\sim$15). Random transition tuples of mini-batch size are sampled from $D$ to train the DQN from past experiences. The target network is used to estimate the target value $y_t$ (Line 13), which is used to evaluate the actions selected by the main Q-network. The loss function is found from (\ref{eq:loss_function}), which is used to update the main Q-network parameters $\theta$ (Line 14). The target network parameters $\theta_{target}$ are updated every fixed number $\tau$ of training steps (Line 15).

Once the training is finished, the proposed DQN learns the placement of the ABS. During application, the trained DQN observes the environment state $s_t$ at each step and selects an action that maximises $f_1(x,y) / f_2(x,y)$. This is repeated multiple times (50 in our case) until the optimum ABS location is found. 

%As for the application of the proposed DQN. Once the training is finished, the proposed DQN learns the placement of the ABS. The DQN observe the environment state $s_t$ at each step. An action is selected to maximise . 

%% algorithm writing reference
% https://shantoroy.com/latex/how-to-write-algorithm-in-latex/

%\begin{figure}[t]
\begin{algorithm}[t]
\DontPrintSemicolon
  \KwInput{Tensor representation of the environment obtained form Algorithm \ref{alg_states}}
  \KwOutput{Final ABS location.}
Initialise the replay memory $D$ to a maximum capacity \\
Initialise Q-network with random weights $\theta$ \\
Initialise target network with weights $\theta_{target}$ = $\theta$ \\
  \For{episode = 1, ..., M} 
    {
        Initialise the environment and receive initial state $s_t$ \\

          \For{step t = 1, ..., T} 
            {
            with probability $\epsilon$ select random action $a_t$\\
            otherwise select $a_t =$ $\max_{a} Q(s_t, a_t;\theta)$\\
            observe $r_t$ and new state $s_{t+1}$ \\
            store transition $\{s_t, a_t, r_t, s_{t+1} \}$ in $D$ \\
        \If{memory is full}
            {sample mini-batch randomly of transitions $\{s_t, a_t, r_t, s_{t+1} \}$ in $D$ \\
            set $y_t = \begin{cases}
                    r_t & \text{$t = T$}\\ 
                    r_t + \gamma \max_{a} Q(s_t, a_t;\theta)  & $\text{$t$} < \text{$T$}$
                        \end{cases} $\;
            update weights for $\theta$ of the main Q-network by minimising the loss function ((\ref{eq:loss_function}))\\
            set $\theta_{target}$ = $\theta$ every $\tau$  steps
            }

}
}
\caption{DRL-based solution for BS placement (training phase)}
\label{alg_DQN_training}
\end{algorithm}
\vspace{-4mm}

%\end{figure}

\begin{figure*}[t]
	\centering
	\subfigure[]{\includegraphics[width = 0.67\columnwidth]{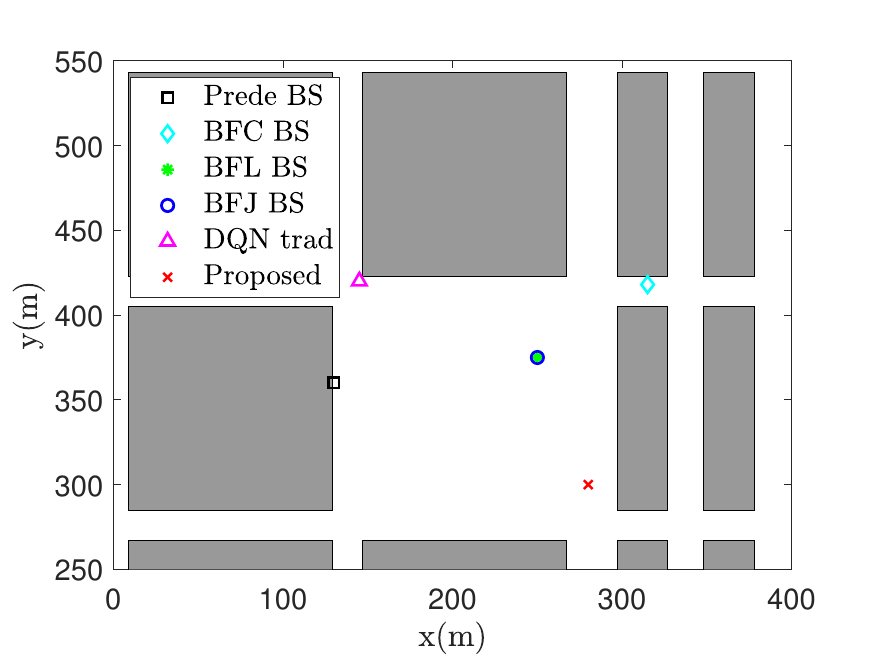}\label{fig:1}}
	\subfigure[]{\includegraphics[width = 0.67\columnwidth]{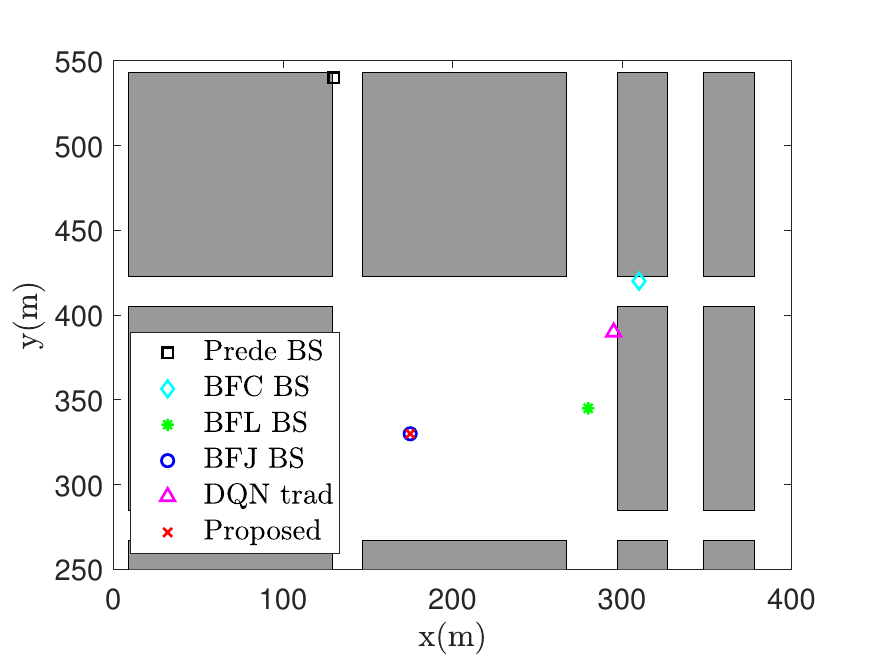}\label{fig:2}}
	\subfigure[]{\includegraphics[width = 0.67\columnwidth]{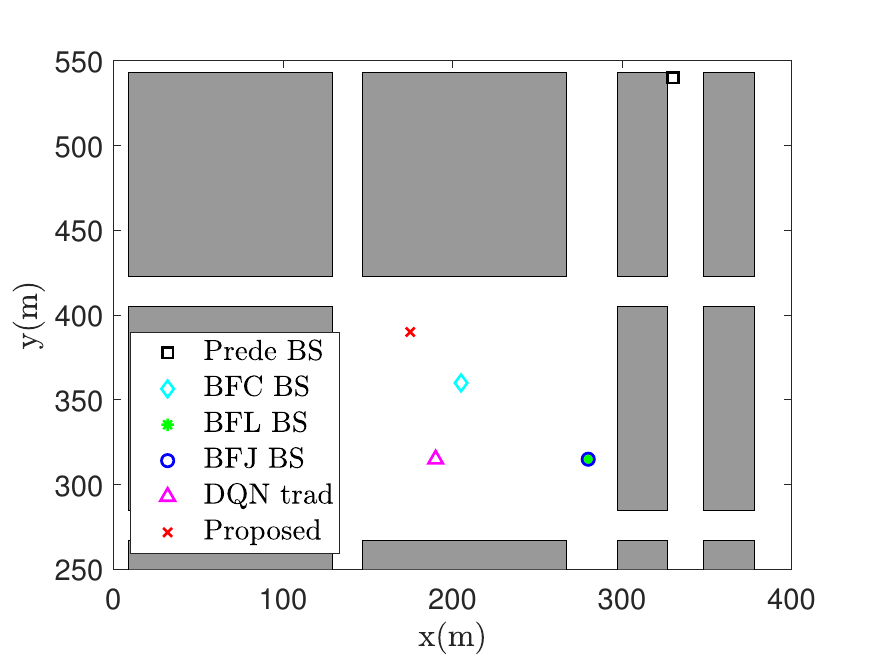}\label{fig:3}}
	\caption{2D visualization for the BS locations. \subref{fig:1} Case 1, \subref{fig:2} Case 2, \subref{fig:3} Case 3.}
	\label{fig:locations_fig_1}	
 \vspace{-2mm}
\end{figure*}

%%%%%%%%%%%%%%%%%%%%

%%%%%%%%%%%%%%%%%%%%
\begin{table*}[]
\centering
\caption{Coverage rate and localisation error (in meter).}
\label{tab:my-table}
\resizebox{2\columnwidth}{!}{%
\begin{tabular}{|c|c|c|c|c|c|c|c|c|c|c|c|c|c|c|c|}
\hline
\textbf{\begin{tabular}[c]{@{}c@{}}Pre-deployed\\ BS\end{tabular}} &
  \multicolumn{3}{c|}{\textbf{BFC}} &
  \multicolumn{3}{c|}{\textbf{BFL}} &
  \multicolumn{3}{c|}{\textbf{BFJ}} &
  \multicolumn{3}{c|}{\textbf{\begin{tabular}[c]{@{}c@{}}Traditional  DQN\end{tabular}}} &
  \multicolumn{3}{c|}{\textbf{\begin{tabular}[c]{@{}c@{}}Proposed DQN\end{tabular}}} \\ \hline
Case & \begin{tabular}[c]{@{}c@{}}Cov rate\\ \end{tabular} & \begin{tabular}[c]{@{}c@{}}Loc error\end{tabular} & \begin{tabular}[c]{@{}c@{}}Cov rate / \\ Loc error\end{tabular} & \begin{tabular}[c]{@{}c@{}}Cov rate\end{tabular} & \begin{tabular}[c]{@{}c@{}}Loc error\end{tabular} & \begin{tabular}[c]{@{}c@{}}Cov rate / \\ Loc error\end{tabular} & \begin{tabular}[c]{@{}c@{}}Cov rate\end{tabular} & \begin{tabular}[c]{@{}c@{}}Loc error\end{tabular} & \begin{tabular}[c]{@{}c@{}}Cov rate / \\ Loc error\end{tabular} & \begin{tabular}[c]{@{}c@{}}Cov rate\end{tabular} & \begin{tabular}[c]{@{}c@{}}Loc error\end{tabular} & \begin{tabular}[c]{@{}c@{}}Cov rate / \\ Loc error\end{tabular} & \begin{tabular}[c]{@{}c@{}}Cov rate\end{tabular} & \begin{tabular}[c]{@{}c@{}}Loc error\end{tabular} & \begin{tabular}[c]{@{}c@{}}Cov rate / \\ Loc error\end{tabular} \\ \hline
1 & 0.98 & 2.45 & 0.40 & 0.90 & 1.46 & 0.61 & 0.90 & 1.46 & 0.61 & 0.88 & 2.19 & 0.40 & 0.92 & 1.73 & 0.53 \\ \hline
2 & 0.95 & 3.37 & 0.28 & 0.86 & 1.97 & 0.43 & 0.87 & 1.97 & 0.44 & 0.90 & 3.21 & 0.28 & 0.87 & 1.97 & 0.44 \\ \hline
3 & 0.98 & 2.17 & 0.45 & 0.95 & 1.71 & 0.55 & 0.95 & 1.71 & 0.55 & 0.95 & 2.35 & 0.40 & 0.98 & 1.87 & 0.52 \\ \hline
\end{tabular}%
}
 \vspace{-2mm}

\end{table*}

\section{Performance analysis}
\label{sec:Performance analysis}
%In this section, we present the numerical results on the localisation accuracy and coverage for the proposed BS placement method. 

\subsection{Evaluation scenario}
% talk about the CNN model, DNN model, table for hyper-parameters for the Q networks

% talk about the training
In the DRL training stage, the DQN model is trained for 3000 episodes with 200 steps per episode. Adam optimiser with adaptive learning rate ($L_r$) was used for the training with $L_r$=$10^{-3}$ for the first 500 episodes, $L_r$=$10^{-4}$  for the following 500 episodes and $L_r$=$10^{-5}$  for the rest of episodes. Reducing $L_r$ allows the optimiser to find the minimum in the loss more efficiently \citep{adavptive_learnign_CVPR}. Parameter $\gamma$ is set to 0.9, the mini-batch size is set to 64, the memory buffer is set to 20000, the target network update frequency $\tau$ is set to 50 and the Mean Squared Error loss function is used. In our study, we compare two DQNs: our proposed version with a grid-based state representation and the traditional version that uses a coordinate-based state representation for both pre-deployed and new Base Stations (BS). The traditional DQN model is structured with a Q-network that consists of two hidden layers containing 50 and 25 neurons, respectively. These layers utilize the ReLU activation function for the hidden layers and a Linear activation function for the output layer. In contrast, our proposed DQN model integrates two CNN layers on top of these hidden layers. The kernel size of these CNN layers is \(4 \times 5\), and there is a \(2 \times 2\) max pooling layer following the first CNN layer. Additionally, the starting location of our ABS is randomly determined at the start of each training episode. To evaluate the model's effectiveness, 70\% of the available pre-deployed BSs are used for training, and the remaining 30\% are utilized for testing. We set the threshold for the received signal strength ($\delta$) to -80 dBm.

%For comparison purposes, we compare our DQN with proposed grid states representation to traditional states representation of coordinates for the predeployed BS and new BS. For traditional DQN, the Q-network included 2 hidden layers with 50, 25 neurons. The ReLU activation function is used for the hidden layers and a Linear activation function is used for the output layer. As for the proposed DQN, we added two CNN layers on top of the aforementioned hidden layers with kernel size $4\times5$ and a max pool $2\times2$ after the first CNN layer. The starting location of our ABS is defined randomly at the beginning of each training episode. Moreover, to verify the results, 70$\%$ of the possible pre-deployed BSs were used for training and 30$\%$ for testing the model. The received signal threshold \delta was set to -80dBm.
% on top of the typical Q-network
% (traditional in terms of state representation as coordinates instead of the proposed grid approach)

\vspace{-1mm}

\subsection{Numerical results}
\vspace{-1mm}

% show that the new states representation can adapt 
% as can be appreciated, the proposed states represntation

% show taining loss for the two algorithms
The achieved localisation error and coverage rate are shown in Table \ref{tab:my-table} for the considered algorithms (i.e., BFC, BFL, BFJ, traditional DQN and proposed DQN) for three different pre-deployed BS scenarios. Fig. \ref{fig:locations_fig_1} shows a 2D visualization of the BS placement for the different approaches. The BSs deployment scenarios order in Table \ref{tab:my-table} matches what is shown in Fig. \ref{fig:locations_fig_1}. As it can be appreciated from Table \ref{tab:my-table}, BFC finds the best location to maximise the coverage rate, but it has a significant impact on the localisation accuracy. Taking Case 2 as an example, the BFC coverage rate is 95\% while the localisation error is more than 3.37 m. On the other hand, the smallest localisation error given by BFL is 1.97 m at the expense of reducing the coverage rate to 86\%. 

%based on our training and reward setting.
The aim of the traditional and proposed DQN algorithm is to provide a solution that is similar to the one obtained from BFJ. Therefore, we also investigate the Cov rate/Loc error ratio. The three shown cases are taken from the test data (locations of pre-deployed BS that are not used while training the DQN). As it can be seen from Table \ref{tab:my-table}, the proposed DQN performance either matches (Case 2) or provides a similar (Cases 1 and 3) coverage and localisation error as BFJ. Therefore, the capability of the proposed DQN to adapt to the changes in the location of the deployed BS is demonstrated, while the traditional DQN approach fails to adapt. The reason why the results for the proposed DQN are not identical to BFJ is as we only trained for 70$\%$ of the possible pre-deployed BSs and the shown results come from the testing (i.e., not seen by the model during training). Moreover, the proposed DQN model needs to be trained only once while BFJ needs to be trained for each scenario. Note that the coverage rate and localisation error achievable by the proposed DQN are not higher than those of BFC and BFL, respectively, as they represent the optimal solutions for coverage rate and localisation error when considered independently. In contrast, the DQN aims for a solution that is jointly optimized.

%Note that the achievable coverage rate and localisation error provided by the proposed DQN is not higher than BFC nor BFL as they represent the optimal solutions independently for coverage rate and localisation error, respectively, whereas DQN pursues a jointly optimised solution.   

% The aim of the traditional and proposed DQN algorithm is to provide a solution that is similar to the one obtained from BFJ. The three shown cases are taken from the test data (locations of pre-deployed BS that are not used while training the DQN). As it can be seen from Table \ref{tab:my-table}, the proposed DQN performance either matches (Case 2) or provides a similar (Cases 1 and 3) coverage and localisation error as BFJ. Therefore, the capability of the proposed DQN to adapt to the changes in the location of the deployed BS is demonstrated, while the typical DQN approach fails to adapt. Note that the achievable coverage rate and localisation accuracy provided by the proposed DQN is not higher than BFC nor BFL as they represent the optimal solutions in terms of coverage rate and localisation accuracy, respectively.  

%significant gains in reducing the localisation error.

%In addition, based on our training and reward setting, we expect that the proposed DQN to approach from the results provided by BFJ. 

%In addition, it is interesting to note that in two scenarios (1 and 3) BFL new BS solution matches BFJ.

%As it can be appreciated, the location of the pre-deployed BS has a significant impact of the final solution (in terms of x and y coordinates of the ABS)

\begin{figure}[!t]
\vspace{-2mm}
	\centering
	\includegraphics[clip, trim=0.0cm 0.cm 0.0cm 0cm, width=1\columnwidth]{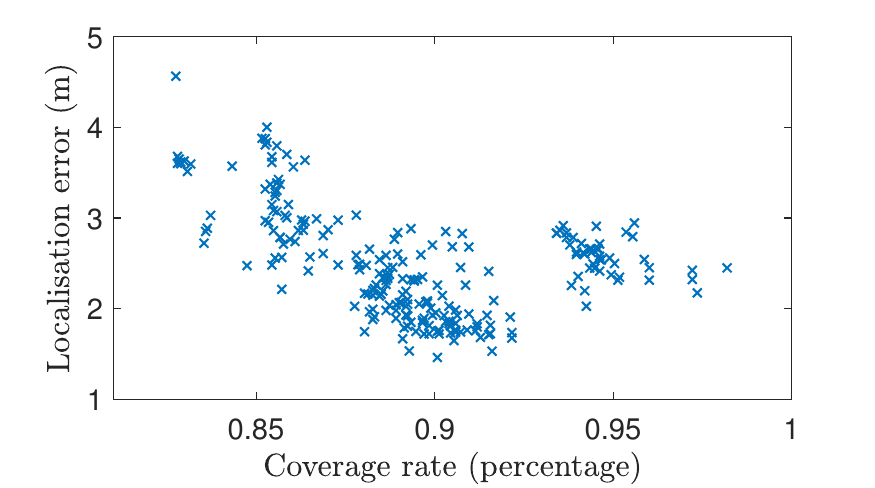}
 \vspace{-6mm}
	\caption{Localisation error vs coverage rate.}
	\label{fig:trade_off1}
\vspace{-4mm}
\end{figure}

Finally, Fig. \ref{fig:trade_off1} shows the effect of selecting different ABS locations and its effect on both localisation error and coverage rate. The pre-deployed BS location is the same as shown in Fig. \ref{fig:locations_fig_1}\subref{fig:1}. The localisation error and coverage rate have different optimal locations and the ABS location to optimise the coverage rate is not the same for optimising the localisation error. 
%In other words, there is a trade-off between the two optimisation problems in the BS location selection which requires to be addressed with network planning according to which parameter is more significant in every particular scenario. 
In other words, there is a trade-off between the two optimisation problems in BS location selection that needs to be addressed through network planning, depending on which parameter is more significant in each specific scenario.

%%%%%%%%%%%%%%%%%%%%%%%%%%%%%%%%%%%%%
% end with conclusions

%\vspace{-1mm}

\section{Conclusions}
\label{sec:Conclusions}
%\vspace{-1mm}

This work has investigated the placement optimisation of mmWave BS in the presence of a pre-deployed BS to simultaneously optimise the coverage rate and localisation accuracy in an urban city layout. We presented a DRL algorithm that is capable to solve the multi-objective problem and adapt to the changes in the location of the pre-deployed BS by proposing a three-layered state representation that is capable to capture spatial properties of the radio environment. Numerical results have demonstrated that the proposed algorithm provides similar results as the optimum exhaustive search algorithms. The reason why the results for the proposed DQN are not identical to BFJ is as we only trained for 70$\%$ of the possible pre-deployed BSs and the shown results come from the testing. Nevertheless, the proposed DQN model needs to be trained only once while BFJ needs to be trained for each scenario. In addition, this work has demonstrated that there is a trade-off between localisation accuracy and coverage rate. Future work will extend the current work to a multi-agent scenario.

\section*{Acknowledgment}
This work was supported by the Academy of Finland Flagship program: Finnish Center for Artificial Intelligence FCAI and the Academy of Finland project 347197 Artificial Intelligence for Urban Low-Emission Autonomous Traffic (AIforLEssAuto).

%\footnotesize
%\bibliographystyle{agsm}
%\bibliography{References}
\footnotesize
\bibliographystyle{IEEEtran}
\bibliography{References}

\end{document}